\documentclass[]{aa}
\usepackage{txfonts}
\usepackage{graphicx}
\usepackage{color}
\usepackage{natbib}
\usepackage{ulem}
\bibpunct{(}{)}{;}{a}{}{,}

\newcommand{\imflux}{\rm erg~s$^{-1}$~cm$^{-2}$}
\newcommand{\kms}{\rm ~km~s$^{-1}$}
\newcommand{\mum}{\rm \,$\mu$m}
\newcommand{\sflux}{\rm erg~s$^{-1}$~cm$^{-2}\ \Box\arcsec^{-2}$~\AA$^{-1}$}
\newcommand{\siflux}{\rm erg~s$^{-1}$~cm$^{-2}$\ \AA$^{-1}\ \Box\arcsec^{-1}$}
\newcommand{\Msun}{M$_\odot$}
\begin{document}

\title{The 3-D Structure of SN 1987A's inner Ejecta\thanks{Based on observations
collected at the European Southern Observatory, Chile (ESO Programme
076.D--0558)}} 
\titlerunning{The Ejecta of SN~1987A Resolved}

\author{Karina Kj\ae r \inst{1,2}
\and Bruno Leibundgut \inst{2,3}
\and Claes Fransson \inst{4,5}
\and Anders Jerkstrand \inst{4,5}
\and Jason Spyromilio \inst{2}}

\institute{Astrophysics Research Centre, Physics Building, Queen's University Belfast, County Antrim, BT7 1NN, United Kingdom\\
\email{k.kjaer@qub.ac.uk}
\and ESO, Karl-Schwarzschild-Strasse 2, D--85748 Garching, Germany
\and Excellence Cluster Universe, Technische Universit\"{a}t
	 M\"{u}nchen, Boltzmannstr. 2, Garching D-85748, Germany
\and Dept. of Astronomy, Stockholm University, AlbaNova, SE-106 91 Stockholm, Sweden
\and The Oskar Klein Centre, Stockholm University
}

\date{Received:  / Accepted }

\abstract
{Observing the inner ejecta of a supernova is possible only in a
handful of nearby supernova remnants. The core-collapse explosion
mechanism has been extensively explored in recent models and 
predict large asymmetries. SN 1987A is the first
modern stellar explosion that has been continuously observed from its beginning to the supernova remnant phase. Twenty years after the explosion, we are now able to observe the
three-dimensional spatially resolved inner ejecta of this supernova.}
{Detailed mapping of newly synthesised material and its radioactive decay daughter products 
sheds light on the explosion mechanism. This may reveal the geometry of the explosion and its
connection to the equatorial ring and the outer rings around SN~1987A.}
{We have used integral field spectroscopy to image the supernova ejecta and
the equatorial ring in the 
emission lines of [Si~I] + [Fe~II] ($\lambda$ 1.64 $\mu$m) and He~I
($\lambda$ 2.058 $\mu$m). The spectral information can be mapped into a 
radial velocity image revealing the expansion of the ejecta both as
projected onto the sky and perpendicular to the
sky plane.}
{The inner ejecta are spatially resolved in a North-South
direction and are clearly asymmetric. Like the ring emission,
the northern parts of the ejecta are
blueshifted, while the material projected to the South of the
supernova centre is moving away from us. We argue that the bulk of the ejecta
is situated in the same plane as defined by the equatorial ring and does not form
a bipolar structure as has been suggested. The exact shape of the ejecta
is modelled and we find that an elongated triaxial ellipsoid fits the
observations best. The velocity measured in the [Si~I] + [Fe~II] line corresponds to 
$\sim 3000$\kms\ and is the same as the width of the IR [Fe~II] line
profiles during the first years. From our spectral analyses of the
ejecta spectrum we find that most of the He I, [Si~I] and [Fe~I-II]
emission originates in the core material which has undergone explosive
nucleosynthesis. The He I emission may be the result of $\alpha$-rich freeze-out if the positron energy is deposited locally.}
{Our observations clearly indicate a non-symmetric explosion mechanism
for SN~1987A. The elongation and velocity asymmetries point towards a
large-scale spatial non-spherical distribution as predicted in recent
explosion models. The orientation of the ejecta in the plane of the
equatorial ring argues against a jet-induced explosion through the
poles due to stellar rotation. } 


\maketitle

\section{Introduction}

Numerical simulations of the core collapse and explosion of massive
stars have shown that, except for progenitors with mass $\la 10
M_{\sun}$ \citep{2006Kitaura}, a one dimensional spherically symmetric
collapse does not produce a successful explosion \citep[e.g.,][]{2006Buras,
2007Burrows, 2006Dessart}. During the last 5-10 years it has, however,
become clear that there are several effects giving rise to different
kinds of multidimensional instabilities and convective motions. The
first of these is the convective motion behind the stalling shock,
induced by the neutrino heating from the proto-neutron star (Bethe \&
Wilson 1985). More recently it has been realised that a large scale
instability, usually known as the standing accretion shock instability
\cite[SASI;][]{2006Blondin}, is a generic feature of the core collapse process,
with most of the power in the $l=1$ mode \citep{2003Blondin}. Several groups have obtained successful explosions by including
these multidimensional effects, although it is too early to draw any definite
conclusions with regard to e.g., the energy of the explosion and the
range of progenitors. 

With this background it is therefore highly interesting to seek as
much observational information about the geometry, kinematics and
abundance structure of the ejecta as possible. There are in the case
of Cas A several indications of large scale mixing and instabilities,
both from the morphology and the elemental abundance distribution
\citep[e.g.,][]{1995Reed,2008Wheeler}. It is, however, likely that Cas
A was a Type IIb SN \citep{2008Krause}, with only a small amount of
hydrogen left, which influences the instabilities and
the general dynamic structure of the ejecta. In this paper we discuss
recent observations of SN~1987A, showing the spatial distribution of
the inner ejecta still powered by radioactive decays. 
This is the first time that it is possible to spatially observe the
evolution of the innermost ejecta as it emerges from the explosion.

SN 1987A is unique in being a very recent explosion, where we have been
able to follow the evolution of the SN from the explosion for more
than 20 years. The fact that we know the mass of the progenitor to be
around $18 M_{\sun}$ \citep{2002Woosley} is particularly important. We
have also accurate estimates of the different isotopic masses, as well
as masses of the most abundant elements \citep[e.g.][]{2002NewFransson}.
From the light curve, emergence of X-rays and line profiles there were
many indications of mixing in the ejecta \citep[e.g.][]{1989Arnett,
1993McCray}. Direct evidence for large scale instabilities came from HST
observations, which have resolved the ejecta, and both the morphology
and the kinematics have been discussed by \cite{2002Wang} where it was 
claimed that the kinematics and morphology as seen in the optical
indicated a bipolar structure, possibly consistent with a jet outflow
and possibly connected to the asymmetry observed of the outer ejecta in
early spectro-polarimetry \citep{1988Cropper, 1991Jeffery, 2008Wang}. The
asymmetry in the explosion was suggested to have the same preferred overall
axis as detected in early speckle interferometric observations 
\citep{1987Meikle, 1987Nisenson} and towards the first emission knot
appearing in the equatorial ring \citep{1998Sonneborn,2002Pun}. The
axisymmetry would include the circumstellar ring structure, the outer
ejecta as well as the inner ejecta \citep{2002Wang, 2008Wang} and was
interpreted as possibly connected to a bipolar or jet-like structure in
the explosion. In this paper we report ground based adaptive optics
observations in the near-IR, which are more naturally interpreted as a
prolate structure in the equatorial plane.

In section \ref{obscal} we describe the observations and calibrations.
In section \ref{res} Figures \ref{contoursfe}, \ref{contourshe}, and
\ref{ylam} visualise the observed spatial distribution of
the ejecta velocity field. We discuss our findings and compare them to
previous observations of SN 1987A in section \ref{disc}, and summarise our
results in section \ref{con}. 

\begin{figure*}
\includegraphics[width=12cm,angle=-90]{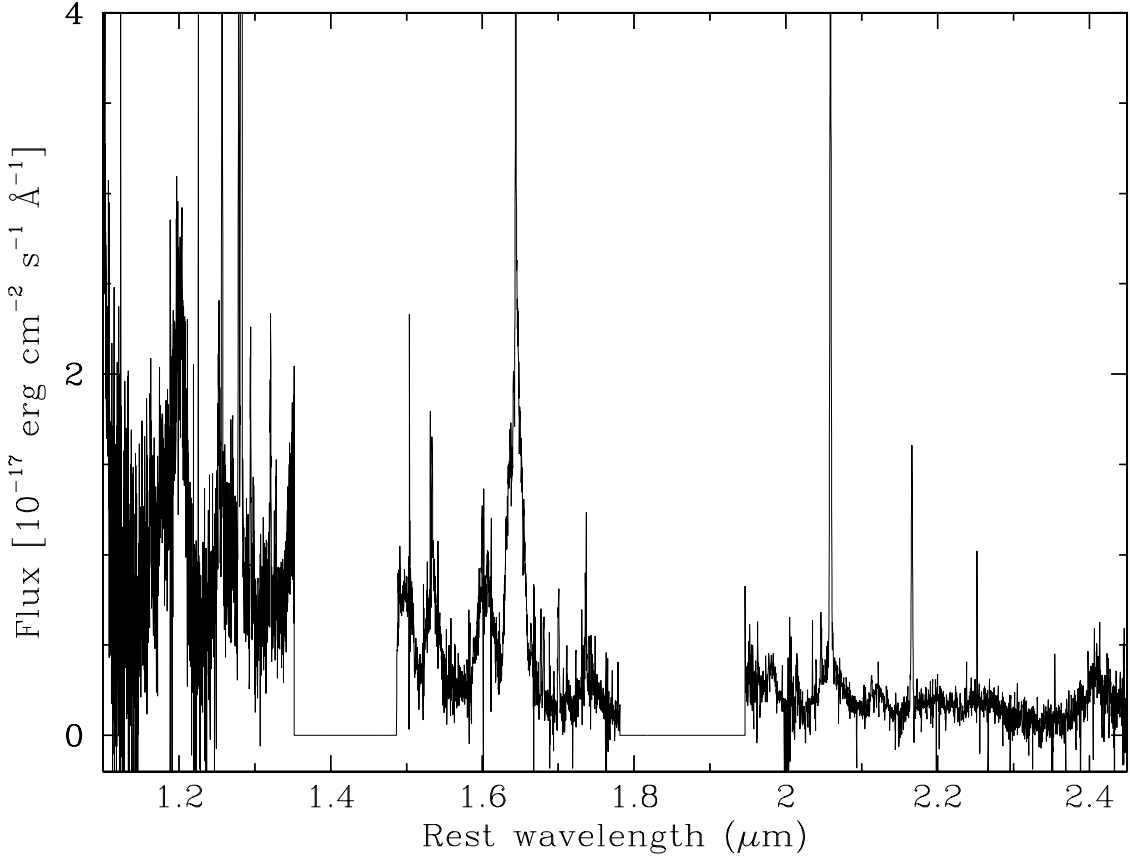}
\caption{Integrated near-IR spectrum of the ejecta of SN~1987A inside a
rectangular box with sides 0.575\arcsec and 0.825\arcsec encompassing
the whole of the ejecta. The integrated area is shown in Fig. \ref{ylam}. }
\label{ejspec}
\end{figure*}

\section{Observations and Calibration}
\label{obscal}
\subsection{Observations}
\label{obs}

Integral Field Spectroscopy observations in the J, H and K
band of the ejecta emission from SN 1987A were
obtained in October and November 2005 with SINFONI
\citep{2003Eisenhauer} on the Very Large
Telescope (VLT) in Chile. The epochs correspond to days 6816, 6824,
6825, 6839, and 6843 since explosion. The observations were supported by
Adaptive Optics (AO) using Star 3 as the reference source. The field of
view (FOV) was $3\arcsec \times 3 \arcsec$ with a spatial resolution of
100$\times$50 mas/pixel for all bands. 

The observations consist of 4800s in J, 4200s in H, and 5400s in K
integrated from single exposures of 600s. All individual exposures were
offset by sub-spaxel spacings to improve the spatial resolution and the
object exposures were separated by sky exposures of equal integration
time in an object-sky, sky-object sequence. The observations were
carried out on 5 separate nights indicated above with airmasses
between 1.4 and 1.5 for all observations.

\subsection{Calibrations}
\label{cal}

The data have been flat fielded, sky subtracted and combined into data
cubes using the SINFONI pipeline version 1.2
\citep{2004ASPCSchreiber,2007pipeline}. Exposures in a filter for a
single night (between 2 and 5 exposures) were combined using the sigma
clipping routine available in the pipeline, where the noise is reduced
by constructing the mean for each data point and omitting data points
that were 2 sigma away from the mean value.

We flux calibrated each night separately using standard stars (HD 76233
(B6V), CCDMJ03187 (G2V), HD 46976 (B9V), HD~52447 (G0V), HD~58112 (B4V),
and HD 94108 (B4V)) and their 2MASS values for magnitudes, types and
colours \citep{2003.2MASS}. In the flux-calibration the observed
standard star spectrum is divided by a template spectrum for that type
of star in order to isolate the instrument response and telluric
features. For the B type standard stars we superimposed hydrogen
absorption features (scaled to the features observed in the standard
stars) on the Planck curve and used that as the stellar template. The
Planck curves are calculated from the theoretical temperatures
corresponding to the stellar type. For the G type stars we used the
solar spectrum scaled with the observed absorption features in the
standard star and brought it to the spectral resolution of the standard
star. 

Data from different nights were combined in order to increase the signal
and to obtain a higher spatial resolution utilising the sub
pixel dithering. The resulting spatial resolution is evident in the
images, where the pixel scale is 25~mas/pixel. 

\begin{table} 
\caption{The Encircled Energy (EE, here the radius) for the H \& K band}
\label{ee-table}
\centering
\begin{tabular}{lcc}
\hline\hline
EE & 50\% & 80\% \\
\hline
H-band & 0.063'' x 0.075'' & 0.150'' x 0.175''\\
K-band & 0.075'' x 0.050'' & 0.150'' x 0.125''\\
\hline
H-band & 2.5 x 3 pixels & 6 x 7 pixels\\
K-band & 3 x 2 pixels & 6 x 5 pixels\\
\hline
\hline
\end{tabular}
\end{table}
We checked the pipeline wavelength calibration against the known
position of the atmospheric OH lines \citep{2000AARousselot} and from
that determined an accuracy of the wavelength calibration as the root
mean square of the correction value. The error of the wavelength
calibration is found to be $1.25\times 10^{-5} \mu$m, $2.1\times
10^{-5}\mu$m and $2.4\times 10^{-5}\mu$m, in J, H and K, respectively.
This corresponds to velocity uncertainties of 3~km~s$^{-1}$,
4~km~s$^{-1}$, and 3~km~s$^{-1}$. 

The spatial resolution of our data is established using the point spread function (PSF) 
of star 3 which mostly lies within our field of view. The detailed knowledge of the PSF 
provides us with a measure of the 
residuals of the adaptive correction of the atmospheric disturbance
and of the spatial resolution of the spectrograph. The PSF has an
enhanced core and broad faint wings and we use the encircled energy metric (EE) to 
quantify the quality of the data.  
Since Star 3 is
not completely sampled in the wings in all bands, we extrapolated the
missing part by using the shape of the PSF by averaging azimuth\-ally.
We measured the EE in the x and y direction independently to spot any differences caused by the FOV image being deconstructed in the y-direction. 
The Encircled Energy (EE) for 80\% and 50\% of the emission from a
point source are summarised in Table \ref{ee-table}. The EEs are shown in some of the images as ellipses (akin to a beam-size) in the upper left corners to display the spread of the emission.

We dereddened the spectra using the galactic extinction law
assuming $R_V=3.1$, and $E_{B-V}$=0.16 \citep{1990AJFitzpatrick} for the
colour excess towards SN~1987A, based on $E_{B-V}$=0.10 from the LMC and
$E_{B-V}$=0.06 from the Milky Way \citep{2003MNRASStaveley}. For the recession velocity we use 286.7 \kms \citep{2008bGroningsson}.

\begin{figure*}
\includegraphics[width=13cm,angle=-90]{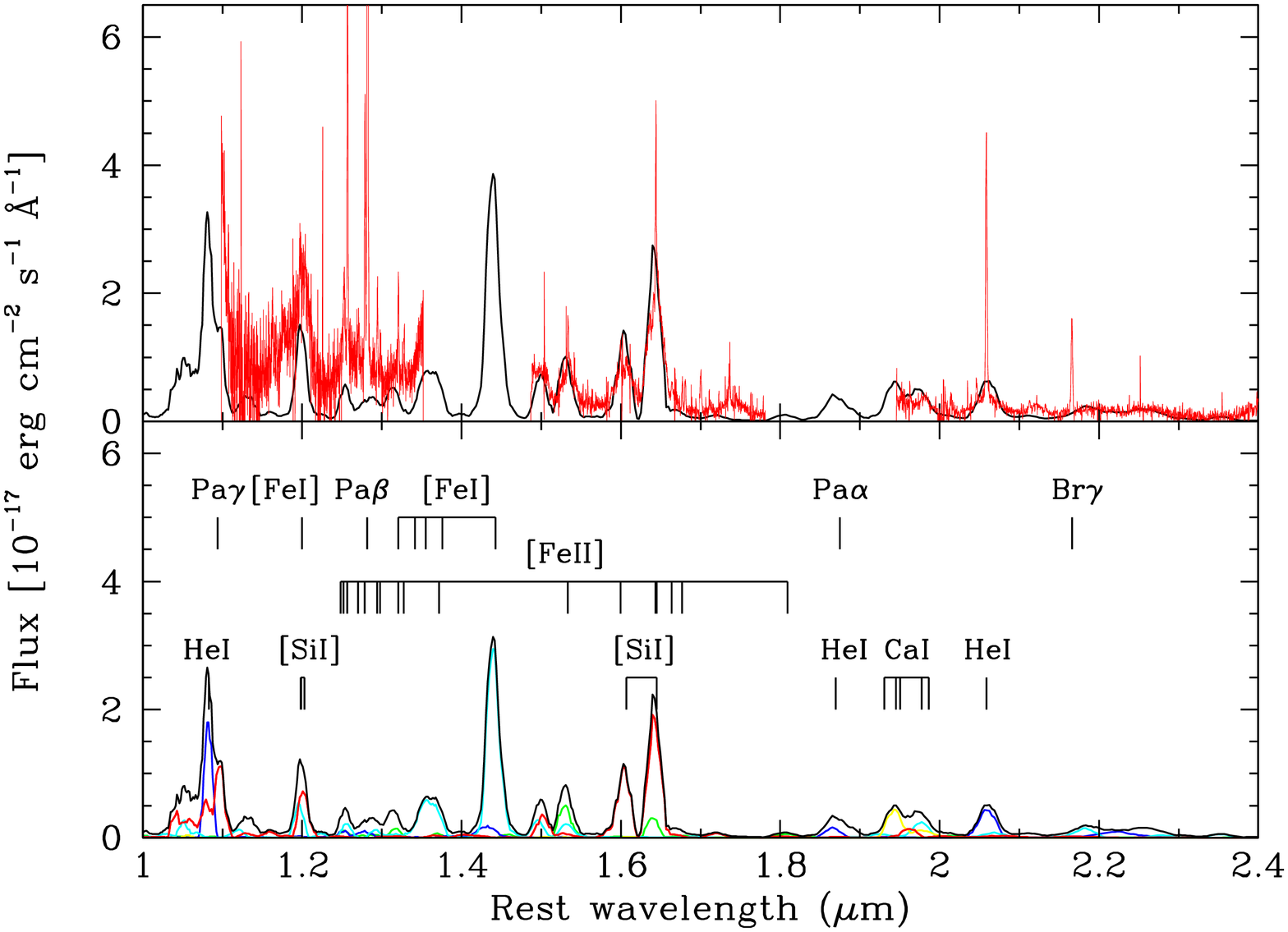}
\caption{
The upper panel shows the observed spectrum together with the total flux from our model calculation. The lower panel shows individual contributions together with line identifications. Dark blue is He I, red is Si~I, yellow is Ca I, cyan is Fe I, green is Fe~II. The narrow lines seen in e.g., Pa$\beta$, [Fe II] 1.644$\mu$m, He I 2.058 $\mu$m and Br$\gamma$ come from scattered light from the ring emission.}
\label{specsimul}
\end{figure*}

\begin{figure*}
\resizebox{\hsize}{!}{\includegraphics[width=17cm]{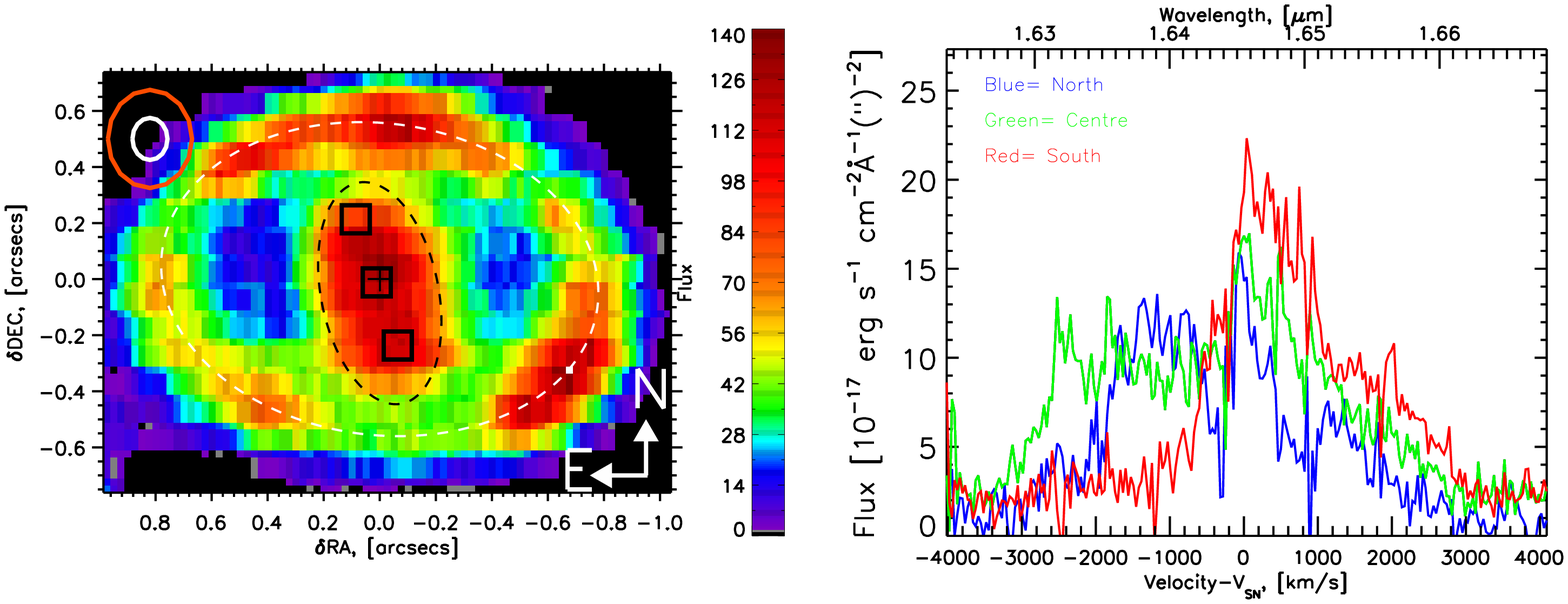}}
\caption{Left panel: image of the 1.644\mum\ [Si~I]+[Fe~II] feature. The white dashed
ellipse indicates the apparent shape of the inner ring, which is centred
on (0,0) marked with a cross. The ejecta shape is indicated by the black
dashed ellipse. The ellipses in the top left corner show the 50\% (80\%
in red) encircled energy area from a point source. The colour bar gives the flux intensity in $10^{-18}$~\imflux. The right
panel shows the line profile of the 1.644\mum\ feature extracted at
three different positions shown in the left panel. The blue curve
corresponds to the upper most extraction box, the green to the middle
box, and the red curve to the bottom box. }
\label{specect}
\end{figure*}

\section{Results}
\label{res}

\subsection{The Ejecta Spectrum}
\label{sec.ejspec}
Our 3D AO spectroscopy allows us to acquire a spectrum of the ejecta and
minimise the contamination from the brighter circumstellar ring. Figure
\ref{ejspec} shows the integrated spectrum
of the ejecta. The integrated area is a
rectangular box with sides 0.575\arcsec and 0.825\arcsec encompassing
the whole of the ejecta (rectangle shown in
Fig.~\ref{ylam}). In addition to the ejecta lines, there are also several narrow
lines coming from the equatorial ring (ER). The contamination of the ejecta 
spectrum by light from the ER arises
from the incomplete correction of the atmosphere by the adaptive optics which, 
as mentioned above, results in a PSF with broad wings. These narrow lines do 
not arise in the ejecta.

From the ejecta we note a number of broad emission lines at 1.535\mum,
1.601\mum, 1.644\mum, and 2.060\mum, already identified in \citet{2007Kjaer}. 
The 2.060\mum\ line is identified as He~I
($\lambda_{\rm rest}=2.058$\mum). The strong lines in the H-band 
are blends of forbidden lines of singly and doubly ionised iron and
singly ionised silicon. We also observe a weak Br$\gamma$ at 2.166\mum.
The region below 1.35\mum\ is more noisy but contains several clear lines at 1.20\mum, 1.25\mum\ and 1.27\mum. Likely identifications of these are discussed below. 

For the later analysis it is important to determine reliable line identifications, especially for the line at 1.644\mum, which 
likely is a blend of [Fe~II] and [Si~I]. Since many of the lines
originate from the same upper levels in Si~I and Fe~II, we made a simple
model by varying the relative populations of those upper levels. We used
Gaussian line profiles with widths of 3000~\kms. These models showed
that the 1.644\mum\ line cannot be pure [Fe~II], since the 1.257\mum~
and 1.322\mum~lines would then be overproduced.
Rather, a mix of both [Si~I] and [Fe~II] is required for a reasonable reproduction of the spectrum.

To obtain more specific line identifications, we have used a
self-consistent model for the spectral formation. This is an
updated version of the code in \cite{1998KozmaI}, which will be
discussed in detail in a forthcoming paper (Jerkstrand, Fransson \&
  Kozma, 2010 in preparation). The main improvement is the addition of a Monte Carlo radiative transfer calculation, allowing for a detailed determination of the internal radiation field. Scattering and fluorescence are taken fully into account, using line opacities from NLTE solutions to the neutral and singly ionised stages of H, He, C, N, O, Ne, Na, Mg, Al, Si, S, Ar, Ca, Sc, Ti, V, Cr, Mn, Fe, Co and Ni. Much of the atomic data base is updated, both for the newly added elements (Sc, Ti, V, Cr, Mn) as well as for the previously included ones. Non-thermal
excitations and ionisations are crucial at this epoch and are included
for all elements along the lines of \cite{Kozma1992}. 

Abundances of the different nuclear burning zones are taken from the 20 $M_\odot$ model in \cite{1995Woosley},
and we assume that these are distributed with individual filling
factors in the core. This prescription mimics the macroscopic mixing
for which there is abundant evidence in SN 1987A
\citep[e.g.][]{1993McCray}. 
At this late phase only $^{44}$Ti is
important as an energy source, the positrons from which dominate
the energy deposition. We assume that these are trapped in the Fe-rich
and Si-rich zones with 90\% energy deposition in the former and 10\%
in the latter, roughly corresponding to the distribution of
$^{44}$Ti in the explosion model. We discuss the sensitivity to the spectrum of this assumption below. The total ${}^{44}$Ti mass used is
$1.0\times 10^{-4}$ \Msun, and we assume a half-life of 58.9 years \citep{Ahmad2006}. The velocity of the core region is 2300 \kms, in rough agreement with the line profiles. We do not make
any assumptions about the composition or location of the dust, but treat it as
a grey absorber within the core. We set an optical depth of $\tau=1$ from the centre to the edge of the core, based
on analyses from earlier epochs \citep[e.g.][]{Lucy1991}. We emphasise that we
  have here not attempted a full investigation of the sensitivity of
  the model to the ${}^{44}$Ti mass, which depends on the assumed
  optical depth of the dust. The ${}^{44}$Ti mass should therefore
  only be taken as indicative. For a determination and extended
  discussion of the ${}^{44}$Ti mass we refer to Jerkstrand, Fransson
  \& Kozma (2010, in preparation).

In Fig. \ref{specsimul} we show the resulting near-IR section of the
spectrum. There are several interesting points to note here. First,
the general agreement is good, given that the model has not been
tweaked to match the data and the free parameters are few. The assumed
${}^{44}$Ti mass should therefore be reasonable, although, as
discussed above, this needs to be confirmed with a full modelling of
the spectrum.  We also confirm that the 1.644\mum\ line is indeed a
mix of [Si~I] and [Fe~II], with [Si~I] being the dominant component ($\sim$80\%). The 1.601\mum\ line is essentially pure [Si~I],
1.535\mum\ is [Fe~I] and [Fe~II] in similar amounts, and the
1.50\mum\ line is a blend of [Si~I] and [Fe~I].

The simulation also shows several
[Fe~I] lines. The strongest of these is the 1.443\mum~line, which
unfortunately lies between the windows in which the atmosphere allows
us to observe. We should, however, point out that the strength of the Fe I lines
depends on uncertain contributions of non-thermal
excitations. Collisional cross sections for these are
largely lacking and are approximated by the Bethe formula.

In the region below 1.35 $\mu$m we identify 1.20 $\mu$m with a blend of
  [Si~I] 1.198,1.199,1.203 $\mu$m and [Fe~I] 1.197 $\mu$m. The line
  at 1.25 $\mu$m is mainly Fe II, and 1.27 $\mu$m is Pa$\beta$. In general, this
  part of the spectrum is too faint in the model. The
  reason might be that the background level in
  the J-band is more uncertain compared to the longer wavelength bands due to the lower Strehl ratio. 
  This means that the wings of the PSF, and
  therefore also the background from scattered light from the ring
  emission, is more important here than in the H- and K-bands. This
  could increase the background considerably in this band. However, the model also contains uncertainties (e.g. the cross sections mentioned above and the positron deposition) that may also explain the
discrepancy.

An interesting result is that the relative strengths between the He~I~2.058\mum\ line and the [Fe II]
lines are well reproduced. At early epochs helium from CNO
nucleosynthesis gives rise to the 2.058\mum\ emission. In our model
most of the helium and iron emission comes from the iron core, although
we cannot exclude some contribution by helium from the hydrogen
burning region, mixed into the core. Evidence for this was seen at
earlier epochs from the line profile of the He I lines
\citep{1998Kozma}, as well as in 2D hydrodynamic simulations
\citep{2006Kifonidis}. The He in the iron core is produced by the
$\alpha$-rich freeze-out. According to the models by \cite{1995Woosley} 
the He abundance in the iron core can be $\sim$50\% by number.
One may in this context note that the He I
2.058\mum\ emission is substantially stronger than the Br$\gamma$ line
(Fig. \ref{ejspec}), which is mainly driven by the recombination
freeze-out, unless positrons leak into this region. The optical depth
to the gamma-rays is small, and only a small fraction of the gamma-ray
energy is absorbed by the ejecta. Recombination freeze-out effects
also affect the helium rich gas from the envelope, and add to the He I
emission from the core. In our model most of the He I emission is,
however, produced by the Fe core. A more quantitative assessment of the
He I emission requires a detailed spectral calculation.

It is not clear whether the positrons are absorbed by the zones containing the ${}^{44}$Ti, where they were emitted, or if they propagate to other zones before being absorbed \citep[see e.g.][]{Chugai1997}. In order to take this uncertainty into account, we explored the consequences of this assumption by computing a model, where we assumed that the positrons were deposited in proportion to the total number of electrons in each zone, rather than at the same location as they were emitted. 

This did not result in any major changes in the model spectrum, although the origin of the emission in some case were different. The 1.601 $\mu$m and 1.644 $\mu$m lines were still dominated by Si I in this model, however now produced by the silicon in the oxygen zones. This alternative model yielded a lower flux for these lines by a factor $\sim$ 2. 
The He I 2.058 $\mu$m line flux did not change much between the two models, but the source of the emission changed from the Fe/He zone to the He and H zones in the alternative model.

\begin{figure*}
\includegraphics[width=18cm]{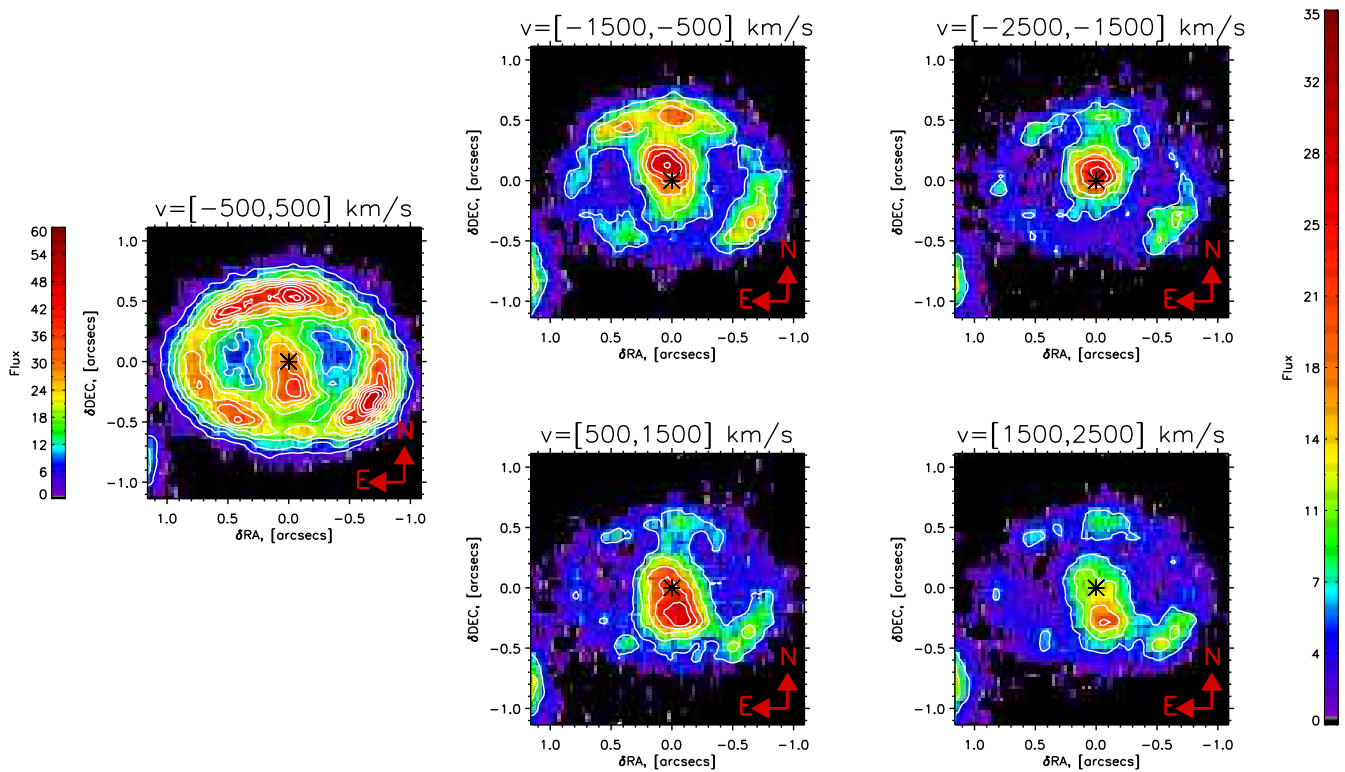}
\caption{
Images of the 1.644\mum\ line of SN~1987A and its circumstellar ring
in different velocity bins along the line of sight, which show the
spatial distribution of the different velocities for that line. The
fluxes are in units of $10^{-18}$~\imflux. The colour bar to the left
indicates the scaling for the middle bin and the colourbar to the
right the remaining bins. The contours trace the intensity levels,
spaced by $5 \times 10^{-18}$~\imflux. The asterisk marks the
centre of the inner ring (as in Fig.~\ref{specect}).  
}
\label{contoursfe}
\end{figure*}

\subsection{The Ejecta Geometry}

We focus in the following on the
1.644\mum\ and He~I~2.058\mum\ lines in our spatial analysis of the
velocity field as, from our data, these lines provide the strongest constraints on the nucleosynthesis.

First we investigate the apparent shape of the ejecta. The left panel of
Fig. \ref{specect} shows an image of the brightest ejecta line at
1.644\mum. The small cross is positioned at the co-ordinates (0,0),
which corresponds to $\alpha=$~05h~35m~28.105s and $\delta~=~$~-69\degr~16\arcmin~10.99\arcsec 
\cite[J2000.0, with the astrometry calibrated with respect to HST images
and the coordinates given by][]{1987West,1993PASPWalborn}. This is the
centre of the white dashed line, that follows the shape of the inner
ring. The shape of the ejecta is elliptical and we find from the
isophotes a ratio of the major to minor axis of $1.8\pm0.17$ (see black
dashed ellipse in left panel of Fig. \ref{specect}). The axis of
symmetry is at a position angle of $15\degr \pm0.9 \degr$, which agrees
quite well with the $14\degr \pm5 \degr$ found from HST observations by
\cite{2002Wang} and \cite{2005ApJSugerman}.

With 3D spectroscopy we have spectral information for each pixel in the
image in the left panel of Fig. \ref{specect}. The right panel of Fig.
\ref{specect} shows the spectra at three different positions indicated
by the squares in the left panel of Fig. \ref{specect}. 
The extraction is performed using an averaging box function 0.1 arcseconds on the side which maximises the signal to noise while not significantly degrading the resolution.
The different spatial positions of the spectra
lead to different line profiles of the 1.644\mum\ line. The
southernmost pixel (red curve) has most red-shifted emission, the
northernmost pixel (blue curve) has most blue-shifted emission, and the
middle pixel (green curve) has both red and blue-shifted emission and
thus a very broad profile. The spectra of the North and South pixels
alone indicate a bipolar structure of the elongated ejecta. The middle
spectrum supports this with its width.

In order to use the spectral information to derive ejecta geometry along
the line of sight we briefly summarise the expected kinematic structure
of freely expanding ejecta. Because of the kinematic nature of the
explosion, once pressure effects have become unimportant, the expansion
is expected to be homologous. With a homologous velocity field, i.e.,
$V(r) \propto r$, constant velocities along the line of sight are
surfaces perpendicular to this direction. The images at the different
wavelengths of an emission line therefore represent a tomography of the
structure of the ejecta along the line of sight. In particular, for a
spherically symmetric ejecta we expect the images at a given velocity
along the line of sight, $V_z$, to be concentric disks with radius $p =
R_0 (1 -(V_z/V_0)^2)^{1/2}$, where $V_0$ is the velocity at the maximum
radius $R_0$. Departures from spherical symmetry will show up as shifts
in intensity in the centre of the images and a non-circular form.

To investigate the asymmetry of the ejecta we show in Figure
\ref{contoursfe} images of the 1.644\mum\ line of SN~1987A in
different velocity intervals. Each frame in the figure shows a separate
part of the ejecta line profile. The divisions are made in velocity,
where the line is integrated over 1000\kms. The upper panels show the
blue-shifted emission and the lower panels show the red-shifted
emission. The contours show the intensity levels spaced by $5~\times~10^{-18}$~\imflux. 

\begin{figure*}
\resizebox{\hsize}{!}{\includegraphics[width=17cm]{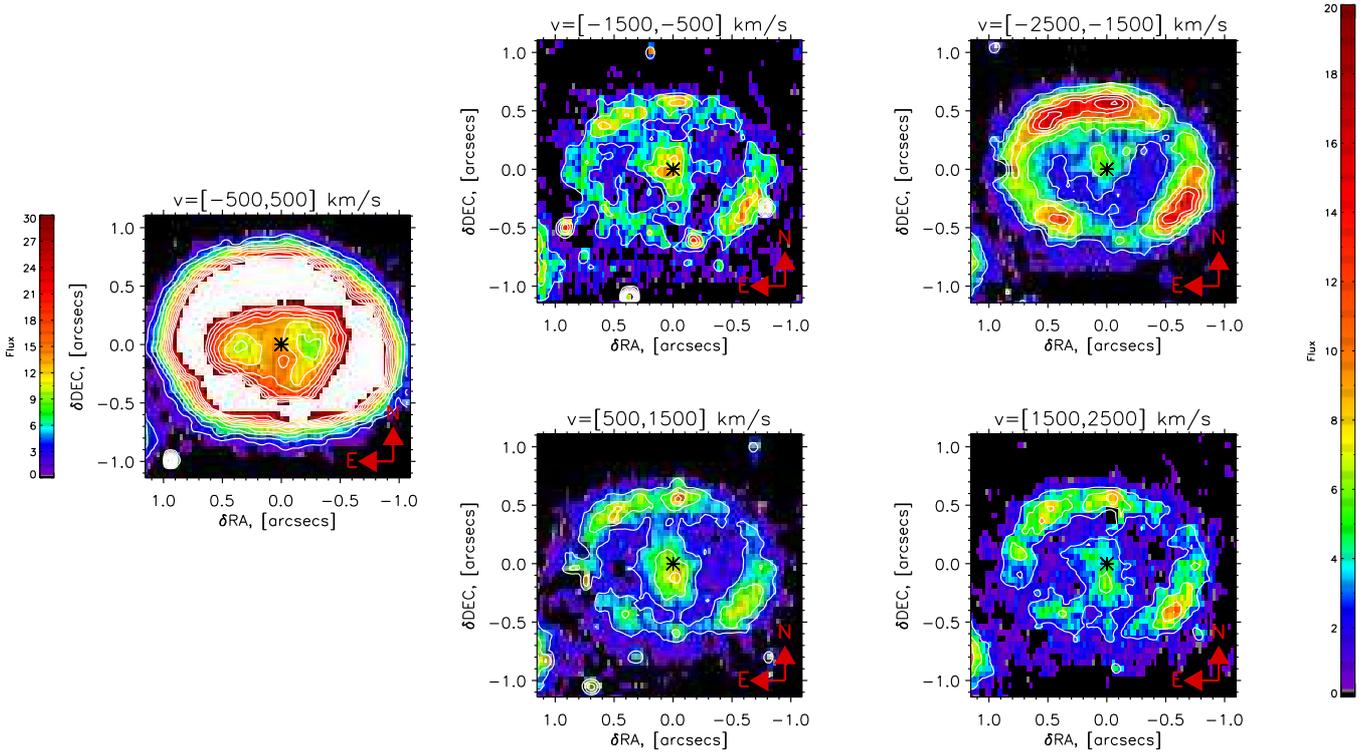}}
\caption{
Images of the He~I~2.058\mum\ line of SN~1987A and its circumstellar
ring in different velocity bins along the line of sight, which show
the spatial distribution of the different velocities for that line.
The units are the same as in Fig.~\ref{contoursfe}.
The contours show the intensity levels, spaced
by $2.5 \times 10^{-18}$~\imflux, and white are saturated areas (to enhance the ejecta against the bright ring).
}
\label{contourshe}
\end{figure*}

The integrated image of He~I~2.058\mum\ clearly shows elongated
ejecta similar to that in the 1.644\mum\ line. We
have compared the 1.644\mum\ and He I images by taking the ratio of
the two, but do not find any significant difference in extent or
intensity distribution. 

Although not the subject of this paper \citep[see
instead][]{2007Kjaer}, we note the dominance of the ring in the
lowest velocity bins of both lines. Furthermore, we also note the stronger emission
from the southern part in the red-shifted $[500,1500]$\kms\ bin, while the
northern part dominates the blue-shifted $[-1500, 500]$\kms\ bin. 
Although the lines from the shocked ring are of intermediate width
(200-400\kms) \citep{2008bGroningsson}, we see that the ring is
visible in all the velocity bins. This is caused by the continuum and
other lines from the ring and/or a broadening of them due to the shock
interaction in the ring. The most important result, which is new, is,
however, that in all velocity bins there is a clear shift in the
position of the maximum intensity between the blue and red sides of the
ejecta, consistent with the velocity pattern seen for the emission from
the ring. This illustrates in a nice way the power of AO-supported
integral field spectroscopy. 

We can also study the geometry in other lines. The spatial distribution
of the kinematics for the He I 2.058\mum\ line is shown in Figure
\ref{contourshe}. Although the emission for the ejecta is much fainter
and the ring is brighter in He I 2.058\mum\ than in the 1.644\mum\ 
line, we observe the same velocity pattern in the emission arising
from the ejecta, with the
blue-shifted emission appearing to the North and the red-shifted component
to the South. The ring emission seen in the [-1500,-2500] \kms~ bin
is most likely due to the [Fe~II] $2.046 \mu$m line \citep{2007Kjaer}.

\subsection{Simulations of the ejecta kinematics}
\label{sec.ejsim}
In order to understand the 3D shape of the ejecta using the spatial
and velocity information in Fig. \ref{contoursfe}, we constructed a
model based on a homogeneous ellipsoid. This is obviously a
simplification in view of e.g., the possible patchy dust obscuration, seen as
a 'hole' in the HST images of the ejecta \citep{2002Wang}, and also
compared to the complex structure seen in realistic simulations
\citep{2006Kifonidis,2009Hammer}. As a first approximation to a non-spherical distribution the
homogeneous ellipsoid provides the necessary constraints for
a qualitative description of the ejecta geometry. 
We have
allowed the axis ratios and the Eulerian
angles of the ellipsoid to vary. Finally, the images have been convolved with
the observed PSF profile for Star 3 in the H-band.

\begin{figure} \resizebox{\hsize}{!}
{\includegraphics[width=17cm]{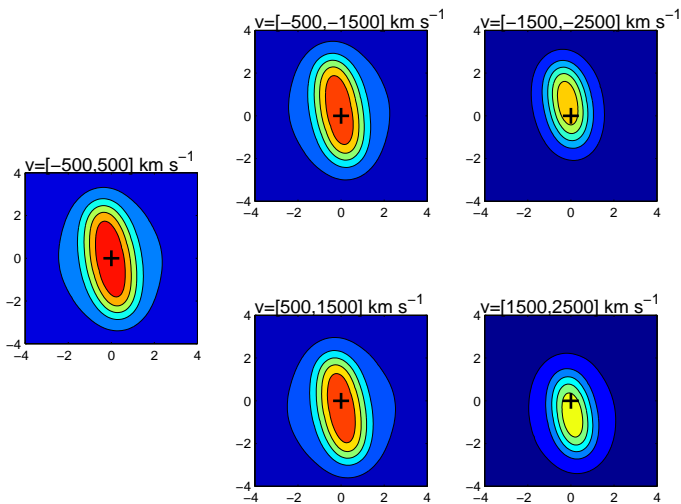}} 
\caption{Simulations of the ejecta intensity from a uniformly emitting
ellipsoid for the same velocity intervals as for Fig.~\ref{contoursfe}. Negative velocities are
in the top row and positive in the bottom. 
The cross marks the centre of the
ejecta. See text for details of the model.} 
\label{sim} 
\end{figure}

Fig. \ref{sim} shows the spatial distribution of the emission in
different velocity bins, scaled to represent that of the observations in
Fig. \ref{contoursfe}. The centroid of the emission is determined by a
combination of the shape of the ejecta, as represented by the axis
ratios, and the orientation in space of the emission region.
Observations during the first years showed strong evidence for dust in the ejecta
\citep{1991Lucy,1993Wooden}. Dust will obscure emission from the far side of the eject (the
red part of the integrated along the line of sight emission). However, since we spatially and spectrally resolve the emission, dust could only impact quantitative measures of the fluxes of lines arising from the ejecta.
The main conclusion of our study, namely, that the ejecta are confined primarily to the equatorial plane, is unlikely to be altered by the effects of dust obscuration. We therefore do not include this effect in the simulations.
 
The best fit to the observations is achieved with
a triaxial body with axis ratios $x:y:z = 3:2:1$, with Eulerian angles
$\theta = 90\degr$ (inclination), $\phi= 25\degr$ (line of nodes) and
$\psi=80\degr$ (angle of major axis). Here we follow the convention for the angles by
\citet{2002clme.book}. 
This
corresponds to a position angle of $10 \degr$ and a tilt out of the
plane of the sky by $25 \degr$. The position angle is consistent within
the errors with that determined directly from the image.

Comparing the model (Fig. \ref{sim}) with the observations we see that
there is a general qualitative agreement between the two as seen 
in Fig. \ref{contoursfe}, with the centroid of emission
indeed shifting from north to south between the blue and red. Also the
axis ratio between the projected images on the plane of the sky
agrees well. Quantitatively there are, however, differences as
expected for this simplified model. While the red-shifted velocity
bins compare reasonably well, the blue-shifted bins have a smaller
spatial extent, corresponding to an ellipsoid, where the major axis
has a larger inclination from the plane of the sky (i.e. viewed more
'head on'). We discuss the ejecta geometry more in section \ref{disc}.

This comparative analysis suggests that the ejecta
shape does not follow a simple ellipsoid. Rather, that the shape of
the blue-shifted part is different from the shape of the red-shifted
part, or at least that the viewing angles of the two are not the
same.

\begin{figure*}
\includegraphics[width=18.5cm]{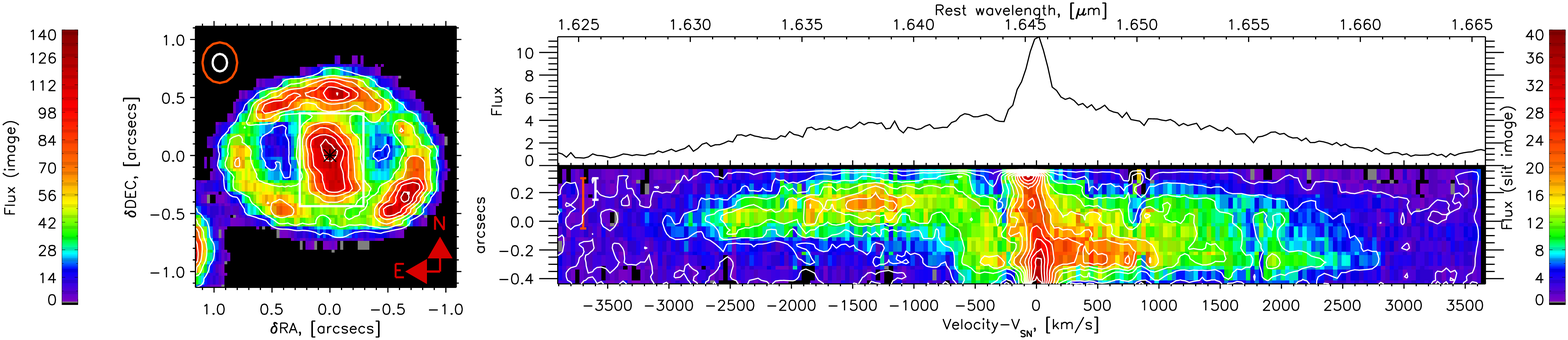}\\
\includegraphics[width=18.5cm]{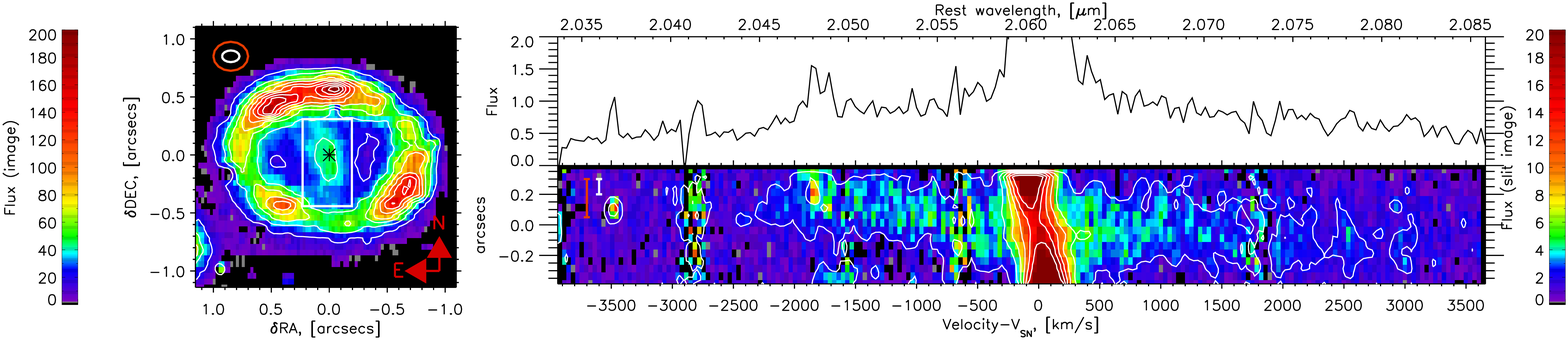}
\caption{Images of the spectral and spatial distribution of the ejecta
lines. Top left: The [Si~I] / [Fe~II] 1.644\mum\ line, integrated over
the spectral range shown in the spectrum to the right. The encircled
energy for 50\% and 80\% of the emission from a star is indicated by
the ellipses. See section \ref{obscal} for details. The colour bar is
in units of $10^{-18}$~\imflux, and the contours are [20, 40, 60, 80, 100, 120, 140] in the same unit. The white box
indicates the integration area for the spectrum shown to the
right. Right: The spectrum, with the bottom panel keeping the spatial
position along the North-South axis, with respect to the asterisk in the
left image. A projection of the encircled energy is shown for 50\% and
80\%. The colour scale follows the colour bar to the right 
with the unit $10^{-18}$~\siflux, and the
contours are [2, 5, 8, 11, 14, .., 41] in the same unit. The
integrated spectrum for the whole area is displayed above, where the
flux is in the unit $10^{-16}$~\sflux. Bottom: Same as above but for
the He~I~2.058\mum\ line.}
\label{ylam}
\end{figure*}

\subsection{The Ejecta Kinematics}
The elongated ejecta visible in images of SN 1987A show 
that the North part of the ejecta is predominately
blue-shifted and the South part is predominately red-shifted.
Figure \ref{ylam} shows the spectral and spatial distribution of the
[Si I]+[Fe II] 1.644\mum\ line and the He I 2.058\mum\ line. The left
panels of the figure are image maps integrated over the spectral range
from the right panels, where the top image is the 1.644\mum\ line
and the bottom image the He I 2.058\mum\ line. The right panels show
the line profiles for the given lines and underneath the spatial
distribution of the emission line profile along the North-South axis,
also referred to as the spectral image. 
The images are centred on the asterisk at coordinates (0,0), which marks the centre of elliptical appearance of the inner
ring. The spectral images follow this convention for the positional
zero point. The Encircled Energy (EE) for 80\% and 50\% of the emission from a
point source are indicated by the ellipses (see Section \ref{obscal} for details).

The projection of the EE ellipses are displayed in the
spectral image, keeping in mind that rather than
displaying a positional error, they display the distribution of the
emission. The white boxes indicate the integration area for the
spectrum shown in the right panels. The area of the boxes are chosen
in order to minimise the contamination from the ring emission while
optimising the emission from the ejecta. The spectral image retains
the North-South axis, showing us the spatial distribution of the red
and blue shifted material. Any information about the spatial
distribution along the East-West axis is lost in this integration. 
Reversing the direction of extraction such that the information in the EW direction is preserved did not result in any additional kinematic information.

The similarity of the spectral image, and thus the kinematics, for the
1.644\mum\ and He~I lines suggests that He is intermixed with the
Fe/Si. This is consistent with an origin of the helium from the
$\alpha$-rich freeze-out in the iron core, as we discussed in
connection to the spectral analysis (section \ref{sec.ejspec}). As was
described there, we can, however, not exclude that mixing of the helium
from hydrogen burning may give the same spatial distribution as the
iron core.

\begin{figure}
\includegraphics[width=9cm]{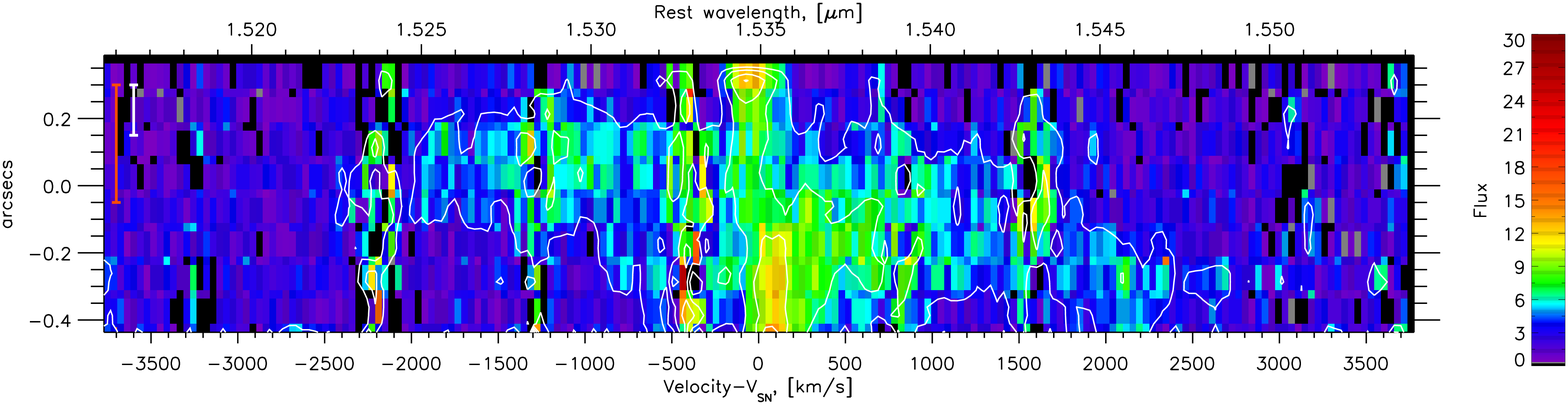}\\
\includegraphics[width=9cm]{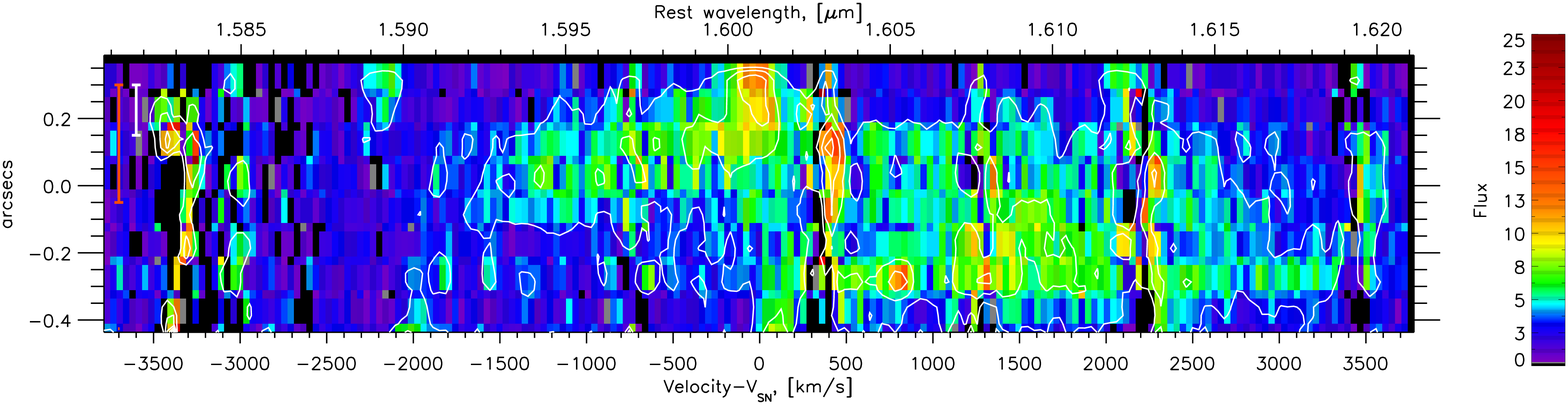}
\caption{Spectral images of the fainter ejecta lines in the H-band,
showing the spatial distribution along the North-South axis. Top: the
[Fe~II]/[Fe~I]~1.53\mum\ line. Bottom: the [Fe II]+[Si~I]~1.60\mum\ line. The
colours follow the colour bars to the right in units of
$10^{-18}$~\siflux\ and the contours trace the intensity levels, spaced by 3 $\times 10^{-18}$~\siflux. The spatial distribution clearly follows the one of
the bright 1.644\mum\ line in Fig.~\ref{ylam}. }
\label{twofe}
\end{figure}

Fig. \ref{twofe} shows the spectral images of two fainter ejecta
lines in the H-band at 1.53\mum\ and 1.60\mum. Here we see
that these lines show the same spatial distribution of the velocities
along the North-South line (y-axis) as the 1.644\mum\ and He~I lines.

Note that we only observe gas which is heated and excited by either
radioactivity or the emission from the shock. The latter is likely to
be important mainly for the outer high velocity ejecta. The inner ejecta
we are discussing in this paper is most likely mainly excited by
the radioactive material. At these late epochs the ejecta is transparent
to the gamma-rays, and the positrons from the ${}^{44}$Ti decay dominate
the excitation. Because these are likely to deposit most of their energy
locally the emission seen from the inner ejecta should mainly reflect
the distribution of radioactive ${}^{44}$Ti. 
The explosion imprinted a velocity structure on the whole ejecta, and it
is this velocity structure that we now see illuminated from within by
radioactive decay.

\section{Discussion}
\label{disc}
\begin{figure}
\resizebox{\hsize}{!}{\includegraphics[width=5cm]{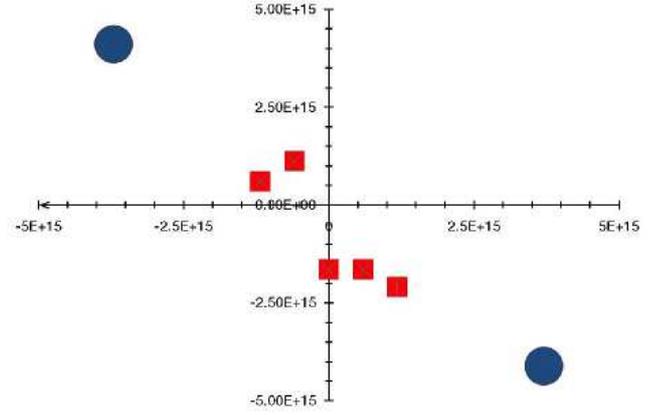}}

\caption{Schematic view of the ejecta distribution of the ejecta
relative to the the ring as seen in the 1.644\mum ~line. Top: The distances from the centre are given in
meters, the observer is located on the left. This figure demonstrates
that the ejecta mostly lie in the same plane as defined by the
equatorial ring. The squares only give approximate emission centres. In reality the emission is more diffuse, see text for details.} 
\label{geo} 
\end{figure}

The observed kinematics and geometry of the ejecta of SN~1987A at late
times clearly indicate that the emission has an orientation generally
similar to the equatorial ring. Fig. \ref{geo} shows the position of
the ejecta emission along the line of sight and compared to the
position of the equatorial ring. We find the inner ejecta of SN~1987A
to be distinctly aspherical, with the blue-shifted material displaced
to the North and the red-shifted material predominately luminous to
the South (Figs. \ref{contoursfe}, \ref{contourshe}, \ref{ylam},
\ref{twofe}). The spatial extent of the red-shifted (southern) material
is significantly larger than that of the blue-shifted material. If we,
in spite of the differences between the blue and red shifted material,
approximate the shape of the whole ejecta with that of an ellipsoid,
we find that the orientation of the major axis is PA=$15\degr$ and
with a tilt out of the plane of the sky of $\sim 25 \degr$. 

The maximum velocities observed are about 2500\kms~ to nearly 3000\kms
(Fig.~\ref{ylam}), corresponding quite well with the velocities
inferred from the line shapes of the radioactive decay lines and the
observed Fe lines in early spectra \citep[e.g.][]{1993McCray}. We
observe material, which is still heated by radioactivity, hence must
come from a region mixed with elements synthesised in the explosion.
With the inner ejecta not perpendicular to the equatorial ring, we find a
configuration, which deviates from the bipolar model proposed
by \cite{2002Wang} and \cite{2008Wang}. 

\citet{2002Wang} discussed the geometry and kinematics of the ejecta
based on HST observations obtained in August 1999, some six years
before our SINFONI observations. The HST image discussed by
\citet{2002Wang} was taken through the F439W filter and is dominated
by emission in [Fe~I] and [Fe~II]. It should therefore be directly
comparable to our [Si~I]+[Fe~II] SINFONI image. The position angle of the
ejecta was found to be $14\degr \pm 5\degr$, consistent with our
results. \citet{2002Wang} also discuss the kinematics based on the
line profile of the [Ca~II] $\lambda~ 7300$ line. By applying
0.1\arcsec slits along the North-South axis, they found that the peak
velocity of the northern part was close to zero, while the southern ejecta has
a positive velocity of $\sim 1700$\kms. These numbers should be
compared to our average velocities of $-1400$\kms and $\sim$700\kms,
respectively (cf. Fig.\ref{specect}). However, as noted by
\cite{2002Wang}, there may be systematic errors in their velocities,
either from the uncertainty in the rest wavelength of the line ([Ca~II] 
$\lambda~7300$ vs. [O~II] $\lambda~7320$), in the wavelength
calibration, or the exact positioning of the slit. Considering these
uncertainties, our measurements may therefore be consistent, {\it if} the
wavelengths by \cite{2002Wang} are blueshifted by $\sim$$1000$\kms. We
note, however, that our spectra have higher S/N, and that we can
trace the line profiles more reliably. In addition, the
ejecta has expanded in the seven years between the two observations
(the supernova age increased by a factor of nearly 1.5) and hence
a better separation of the different emission regions can be achieved.
Finally, thanks to the integral field spectroscopy, we can image the
ejecta emission over the whole ejecta, as well as in specific lines, which represent a 
considerable advantage compared to the HST filter observations.

The observed geometry of the inner ejecta strongly supports a
large-scale instability, like SASI, in the explosion. The strong
mixing of material and the asymmetry of the explosion in SN~1987A, as
already indicated by the bolometric light curve, the Bochum event,
line profiles and the early emergence of $\gamma-$ and X--rays is here
clearly confirmed. The kinematics of the inner ejecta show that the
explosion has defined an orientation, which is close to the
equatorial plane of the progenitor star as defined
by the circumstellar rings. This would invalidate jets produced
through the poles, if the equatorial ring really defines a rotation
axis of the progenitor star. 

In addition, the inner ejecta appear not to be located in a plane, but
rather we see two extensions at different radial velocities and
projected spatial extent. The strongest emission in the northern
extension is located at about $-$1400\kms and at 0.15\arcsec from the
explosion centre, while the southern part displays the strongest
emission at a velocity of $\sim$500\kms but extends out to nearly
0.3\arcsec. Fig.~\ref{geo} displays the situation of the ejecta
relative to the equatorial ring. The ejecta velocities have been
converted to the line-of-sight positions at the time of the
observations (6840 days) and the angular distance in declination
($\delta$DEC) assuming a distance to SN~1987A of 50 kpc. The ring
position was determined from measured offset from the centre and
assuming an inclination of $41 \degr$ \citep{2005ApJSugerman}. This
figure should be seen as a more detailed 2D complement to
Fig. \ref{sim}, which, however, gives a full 3-D representation,
although with the approximation of an ellipsoid. The general
orientation of the major axis is consistent within the errors of these
two representations.

Since the inner ejecta seem not to be co-planar, we may also speculate
about the direction of the kick received by the neutron star.
Presumably, the inner ejecta are the part, which impart the momentum
of the large hydrodynamic instabilities onto the proto-neutron star.
The connection between the SASI instabilities on scales of a
few$\times 10^3$ km and the large scale asymmetry we see is not
obvious. The first calculations by \cite{2006Kifonidis} and
\cite{2009Hammer} show that the SASI is important for triggering of
the instabilities, although it is not clear how much of the initial
instability survives the dynamics of the explosion on the much longer
time scale of hours and days until the hydrodynamic structure is
frozen-in. It is therefore too early to speculate about a specific
direction for the kick of the neutron star. It should be recalled that the images reflect the distribution of the positron input and although the positrons are stipulated to have a limited travelling range their distribution cannot be taken to be the distribution of mass in the core.

As was discussed in Sect. \ref{sec.ejsim}, dust affects the
  observed flux from especially the red side of the emission. 
  The limited signal-to-noise and the fact that most lines are blended make a detailed analysis of
  the dust difficult. The fact
  that we do observe substantial emission from red-shifted velocities (e.g.,
  Figs. \ref{specect} and \ref{ylam}) shows that the dust can only
  block the core partially at most, as it has earlier in the evolution. 
  Even so, the line profiles of the rather clean [Si I] 1.60 and 1.64 $\mu$m
  lines show clear deficiencies of the red sides.
  We have not attempted any investigation to see whether these asymmetries
  have quantitatively changed compared to earlier epochs \citep[e.g.][]{2002Fassia}, mainly because of the limited resolution and signal-to-noise of these observations.
 
\section{Conclusions}
\label{con}

The outer ejecta of SN~1987A travelling at around 30000\kms\ have reached
the outer ring a decade ago, and have become visible mostly in the
reverse shocks generated by the collision with the ring material
\citep{2003McCray, 2006Groningsson, 2008aGroningsson, 2008bGroningsson,
2007Kjaer, 2010Zanardo}. The inner ejecta are still heated by the
radioactive decays \citep[mostly positrons from the $^{44}$Ti
decays;][]{2002NewFransson} and have expanded enough to be spatially
resolved. The extent of this material corresponds to about 3000\kms. Our spectral analysis shows that most of the helium emission comes from the helium produced in the $\alpha$-rich freeze-out in the core. This is also supported by the similar spatial distribution of the He I and [Si I]+[Fe II] images. 
Helium zone material mixed into the core may, however, also contribute to the He I emission.

We have confirmed the asymmetric shape of the inner ejecta in SN~1987A
and have shown that it is confined roughly to a similar plane as the
equatorial ring. Also, the northern and southern lobes are not
symmetric and show slightly different radial velocities, which points towards
two separate angles in the line of sight of the emission sites. Both
these arguments are against a jet-induced explosion as favoured by
\citet{2002Wang} due to an explosion propagating through the poles of
a rotating star. Instead, the shape of the inner ejecta are fully
consistent with what is expected from the large instabilities
predicted in recent explosion models of core-collapse supernovae. The
SINFONI observations of SN~1987A can be seen as a direct
observational confirmation of these models. 

\begin{acknowledgements}
We would like to thank the Garching and Paranal Astronomers who
provided support during the service observations with SINFONI. We are
also grateful to Thomas Janka, Cecilia Kozma, Dick McCray and
Craig Wheeler for discussions. This work was supported by the Swedish
Research Council and the Swedish National Space Board (CF, AJ, KK). KK has
been supported by a Carlsberg Foundation Fellowship and by Queen's
University Belfast, Northern Ireland.
\end{acknowledgements}

\bibliographystyle{aa}
\bibliography{14538}

\end{document}